\documentclass[final]{aipproc}

\newcommand{\aislimit}{m$_{AB}\simeq 20.5$}
\newcommand{\mislimit}{m$_{AB}\simeq 23$}
\newcommand{\dislimit}{m$_{AB}\simeq 25$}
\newcommand{\sqdeg}{deg$^2$}

\newcommand{\ha}{\,H$\alpha$} 
\newcommand{\hb}{\,H$\beta$}

\layoutstyle{6x9}

\SetInternalRegister\hbadness{8000} 

\begin{document}

\title 
      [GALEX: Galaxy Evolution Explorer]
      {GALEX: Galaxy Evolution Explorer}


\author{Barry F. Madore}{
  address={Carnegie Observatories, Pasadena CA 91101,
  e-mail: barry@ociw.edu}
}

\copyrightyear  {2005}

\begin{abstract}
We review recent scientific results from the Galaxy Evolution Explorer
with special emphasis on star formation in nearby resolved galaxies.
\end{abstract}

\date{\today}

\maketitle

\section{Introduction}

\subsection{The Satellite}

The Galaxy Evolution Explorer (GALEX) is a NASA Small Explorer class
mission. It consists of a 50~cm-diameter, modified Ritchey-Chr\'etien
telescope with four operating modes: Far-UV (FUV) and Near-UV (NUV)
imaging, and FUV and NUV spectroscopy. The telescope has a 3-m focal
length and has Al-MgF$_2$ coatings.  The field of view is
1.2$^{\circ}$\ circular.  An optics wheel can position a CaF$_2$
imaging window, a CaF$_2$ transmission grism, or a fully opaque mask
in the beam.  Spectroscopic observations are obtained at multiple
grism-sky dispersion angles, so as to mitigate spectral overlap
effects.  The FUV (1528\AA: 1344-1786\AA) and NUV (2271\AA:
1771-2831\AA) imagers can be operated one at a time or simultaneously
using a dichroic beam splitter. The detector system encorporates
sealed-tube microchannel-plate detectors. The FUV detector is preceded
by a blue-edge filter that blocks the night-side airglow lines of
OI1304, 1356, and Ly$\alpha$. The NUV detector is preceded by a red
blocking filter/fold mirror, which produces a sharper long-wavelength
cutoff than the detector CsTe photocathode and thereby reduces both
zodiacal light background and optical contamination.  The peak quantum
efficiency of the detector is 12\% (FUV) and 8\% (NUV). The detectors
are linear up to a local (stellar) count-rate of 100 (FUV), 400 (NUV)
cps, which corresponds to m$_{AB}\sim 14-15$.  The resolution of the
system is typically 4.5/6.0 (FUV/NUV) arcseconds (FWHM), and varies by
$\sim$20\% over the field of view. Further detail about the mission,
in general, and the performance of the satellite, in specific, can be
found in Martin {\it et al.}  \cite{mar05} and Morrissey {\it et al.}
\cite{mor05}, respectively. The mission is nominally expected to last 38
months; GALEX was launched into a 700~km, 29$^{\circ}$\ inclination,
circular orbit on 28 April 2003.

\vfill\eject

\subsection{The Prime Mission}

GALEX is currently undertaking the first space UV sky-survey,
including both imaging and grism surveys. The prime mission includes
an all-sky imaging survey (AIS: 75-95\% of the observable sky, subject
to bright-star and diffuse Galactic background light limits)
(\aislimit), a medium imaging survey (MIS) of 1000 \sqdeg (\mislimit),
a deep imaging survey (DIS) of 100 square degrees (\dislimit), and a
nearby galaxy survey (NGS). Spectroscopic (slitless) grism surveys
(R=100-200) are also being undertaken with various depths and sky
coverage. Many of the GALEX fields overlap existing and/or planned
ground--based and space-based surveys being undertaken in other bands.

{\it All-sky Imaging Survey (AIS):} The goal of the AIS is to survey
the entire sky subject to a sensitivity of \aislimit, comparable to
the POSS II (m$_{AB}$=21~mag) and SDSS spectroscopic (m$_{AB}$=17.6~mag)
limits. Several hundred to 1,000 objects are in each 1
\sqdeg~field. The AIS is performed in roughly ten 100-second pointed
exposures per eclipse ($\sim$10 \sqdeg~per eclipse).

{\it Medium Imaging Survey (MIS):} The MIS covers 1000 \sqdeg, with
extensive overlap of the Sloan Digital Sky Survey.  MIS exposures are
a single eclipse, typically 1500 seconds, with sensitivity \mislimit,
net several thousand objects, and are well-matched to SDSS photometric
limits.

Bianchi {\it et al.} \cite{bia05a} have made the first characterization of
the objects detected in the GALEX AIS and DIS surveys by
cross-comparing objects already discovered and classified on the Sloan
Digital Sky Survey in regions of overlap. Emphasis was put on
discovering quasars and for the 75 and 92 square degrees studied in
the MIS and AIS overlap with SDSS they report 1,736 and 222 z $\le$ 2
QSO candidates, for the two GALEX survey depths, respectively.  This
significantly increases the number of fainter candidates, and
moderately augments the candidate list for brighter objects.

\begin{table}
\begin{tabular}{lcc}
\hline
\tablehead{1}{l}{b}{Item}
&\tablehead{1}{c}{b}{FUV Band}
&\tablehead{1}{c}{b}{NUV Band}  \\
\hline
Bandwidth:    & 1344~--~1786~\AA             & 1771~--~2831~\AA \\
Effective wavelength ($\lambda_{\rm eff}$):
                               & 1528~\AA                     & 2271~\AA\\
Field of view:                 & 1.28$^{\circ}$               & 1.24$^{\circ}$\\
Zero point ($m_{\rm 0}$):      & 18.82~mag                    & 20.08~mag\\
Image resolution (FWHM):       & 4.5~arcsec                   & 6.0~arcsec\\
Spectral resolution ($\lambda/\Delta\lambda$):
                               & 200                          & 90 \\
Detector background (typical):\\
\hspace{2em}Total:             & 78 c-sec$^{-1}$              & 193 c-sec$^{-1}$\\
\hspace{2em}Diffuse:           & 0.66 c-sec$^{-1}$-cm$^{-2}$  & 1.82 c-sec$^{-1}$-cm$^{-2}$\\
\hspace{2em}Hotspots:          & 47 c-sec$^{-1}$              & 107 c-sec$^{-1}$\\
Sky background (typical): 
                               & 2000 c-sec$^{-1}$            & 20000 c-sec$^{-1}$\\
Limiting magnitude ($5\sigma$):\\
\hspace{2em}AIS (100~sec):       & 19.9~mag                   & 20.8~mag \\
\hspace{2em}MIS (1500~sec):      & 22.6~mag                   & 22.7~mag \\
\hspace{2em}DIS (30000~sec):     & 24.8~mag                   & 24.4~mag \\      \hline
\end{tabular}
\caption{Selected Performance Parameters (Morrissey {\it et al.} 2005)}
\label{tab:a}
\end{table}

\vfill\eject

{\it Deep Imaging Survey (DIS):} The DIS consists of 20 orbit (30
ksec, \dislimit) exposures, over 80 \sqdeg, located in regions where
major multiwavelength efforts are already underway.  DIS regions have
low extinction, low zodiacal and diffuse galactic backgrounds,
contiguous pointings of 10 \sqdeg~ to obtain large cosmic volumes, and
minimal bright stars. An Ultra DIS of 200 ksec, m$_{AB}\sim 26$~mag is also
in progress in four fields.

{\it Nearby Galaxies Survey (NGS):} The NGS targets well-resolved
nearby galaxies for 1-2 eclipses.  Surface brightness limits are
m$_{AB}\sim$27.5~mag arcsec$^{-2}$.  The 200 targets are a diverse
selection of galaxy types and environments, and include most galaxies
from the Spitzer IR Nearby Galaxy Survey (SINGS).  Figure 1 shows the
NGS observation of the M81/M82 system.

{\it Spectroscopic Surveys.}  The suite of spectroscopic surveys
includes (1) the Wide-field Spectroscopic Survey (WSS), which covers
the full 80 \sqdeg~DIS footprint with comparable exposure time (30
ksec), and reaches m$_{AB}\sim 20$~mag for S/N$\sim$10 spectra; (2) the
Medium Spectroscopic Survey (MSS), which covers the high priority
central field in each DIS survey region (total 8 \sqdeg) to
m$_{AB}$=21.5-23.0~mag, using 300 ksec exposures; and (3) Deep Spectroscopic
Survey (DSS) covering 2 \sqdeg~with 1,000 eclipses, to a depth o f
m$_{AB}$=23-24~mag.

\section{NEARBY GALAXY SURVEY (NGS)}

\subsection{NGC 0253 and M82}

\begin{figure}
  \includegraphics[height=.7\textheight]{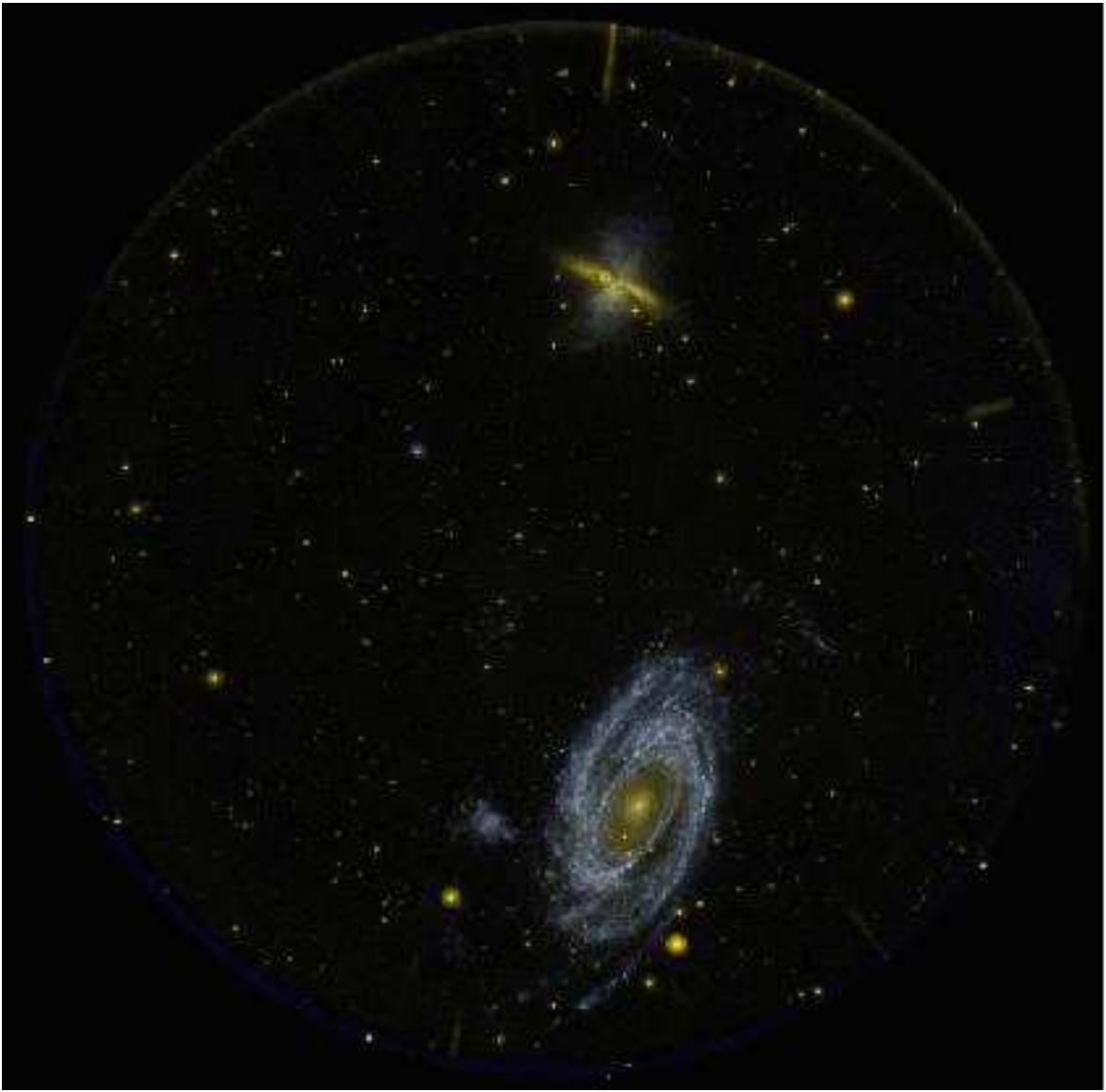} \caption{GALEX
  Composite NUV/FUV Image of M81 (bottom) and M82 (top). Note the
  verey different morphology of M82 as compared to its optical
  image. The scattering of light by dust ejected by the starburst is
  especially prominent in the blue light detected by GALEX. On the
  other hand, the bulge of M81 is almost invisible in the GALEX bands,
  leaving only the complex array of star-forming filaments detailing
  the spiral structure of this galaxy. The irregular object to the
  east of M81 is Holmberg IX. Another interacting galaxy in the
  M81-M82 system, NGC 3077 is outside the field of view to the east
  and may never be imaged by GALEX because of a nearby bright star. }
\end{figure}

In a study of the complex of ultraviolet filaments in the
starburst-driven halos of NGC 0253 and M82 Hoopes {\it et al.} \cite{hoo05}
concluded that the UV luminosities in the halo are too high to be
provided by continuum and line emission from shock-heated or
photoionized gas except perhaps in the brightest filaments in M82.
This suggests that most of the UV light is the stellar continuum of
the starburst scattered into our line of sight by dust in the
outflow. This interpretation agrees with previous results from optical
imaging polarimetry in M82. The observed luminosity of the halo UV
light is $\le 0.1\%$ of the bolometric luminosity of the
starburst. The morphology of the UV filaments in both galaxies shows a
high degree of spatial correlation with \ha\ and X-ray emission, which
indicates that these outflows contain cold gas and dust, some of which
may be vented into the intergalactic medium (IGM). UV light is seen in
the ``\ha\ cap'' 11 kpc north of M82. If this cap is a result of the
wind fluid running into a pre-existing gas cloud, the gas cloud
contains dust and is not primordial in nature but was probably
stripped from M82 or M81 (Figure 2).  If starburst winds efficiently
expel dust into the IGM, this could have significant consequences for
the observation of cosmologically distant objects.
\vfill\eject

\begin{figure}
\caption{GALEX Montage of M31. As large as the field of view of GALEX
is, it still requires many pointings (in this case 9 images) to cover
the main body of the galaxy and part of its halo. Both M32 and NGC 205
are visible in this image, but because of their intrinsically red
population they are not as conspicuous as they are in the visible.}
\includegraphics[height=.7\textheight]{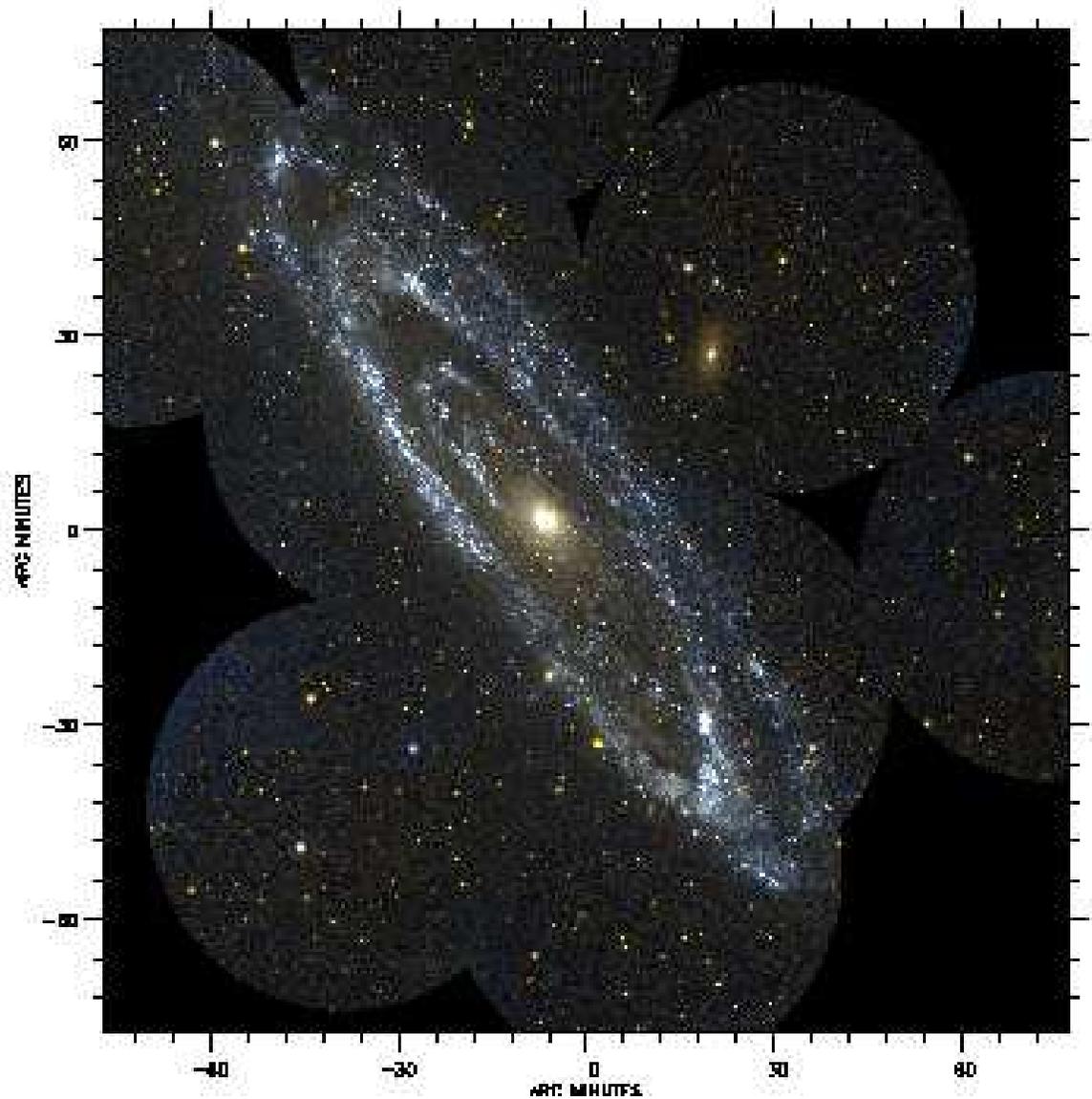}
\end{figure}

\subsection{M31, M32 and M33: Radial Gradients}

For M31 and M33 (Figures 3 \& 4, respectively) Thilker {\it et al.}
\cite{thi05a} have determined the UV radial light and color profiles and
compared them with the distribution of ionized gas as traced by
H$\alpha$ emission.  It is found that the extent of the UV emission,
in both targets, is greater than the extent of the observed $HII$
regions and diffuse ionized gas.  In addition, the ultraviolet diffuse
fraction in M33 using the FUV observations has been compared to the
H$\alpha$ diffuse fraction obtained from wide-field narrow-band
imaging.  The FUV diffuse fraction appears to be remarkably constant
near 0.65 over a large range in galactocentric radius, with departures
to higher values in circumnuclear regions and, most notably, at the
limit of the H$\alpha$ disk. The authors suggest that the increase in
FUV diffuse fraction at large galactocentric radii could indicate that
a substantial portion of the diffuse emission beyond this point is not
generated {\it in situ} but rather it is scattered from dust, after
originating in the vicinity of the disk's outermost $HII$ regions. A
radial variation of the H$\alpha$ diffuse fraction was also measured:
in general the H$\alpha$ diffuse fraction is near 0.4 but it rises up
to 0.6 toward the galaxy center.

\begin{figure}
\includegraphics[height=.7\textheight]{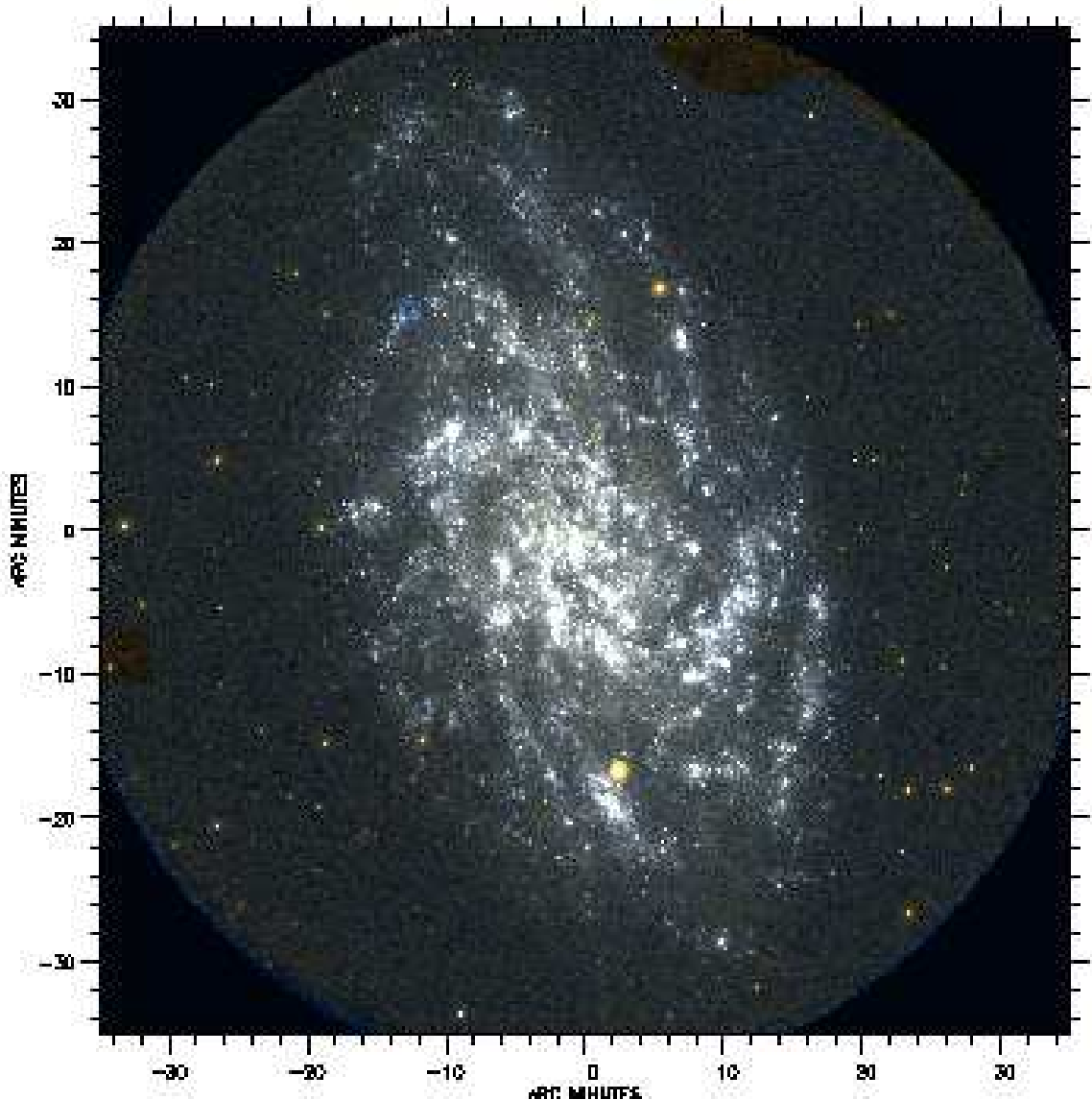}
\caption{GALEX Montage of M33. A single pointing of GALEX is capable
of imaging almost the entire main body of M33. Because M33 has only a
faint background population of old red stars its morphology in the UV
is not too dissimilar to its appearance at the optical
wavelengths. The primary difference being the contrast of the arms with
respect to the disk.}
\end{figure}

For M32, the high-surface-brightness elliptical companion
to the Andromeda galaxy, Gil de Paz {\it et al.} \cite{gdp05} have
undertaken a detailed analysis of its UV light and color
profile. Early studies of the UV properties of this galaxy, first by
O'Connell {\it et al.} \cite{con92} and then by Ohl {\it et al.} 
\cite{ohl98} cast some
doubt on M32 as being a truly typical elliptical galaxy. Based on UIT
(Ultraviolet Imaging Telescope) data it was claimed that M32 had a
strong FUV-optical color gradient, but that it was inverted (negative)
with respect to the gradients observed in the vast majority of other
elliptical galaxies. If true, M32 would be extremely anomalous. The
imaging data for M32 had to be carefully decomposed so as to properly
account for the complicated background contamination by the disk of
M31. The results are shown in Figure 5 where it can be seen that M32
has a very modest (positive) color gradient making M32 fully
consistent in its UV properties as compared to those observed in
luminous ellipticals.

\begin{figure}
  \includegraphics[height=.3\textheight]{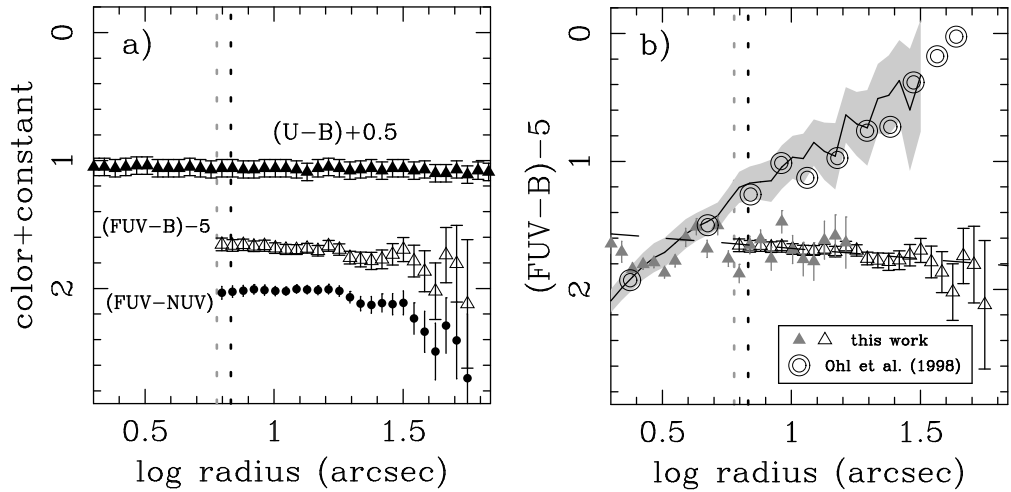} \caption{In Gil
  de Paz et al. \cite{gdp05} the luminosity profiles and color gradients
  have been derived for the compact elliptical galaxy M32, a nearby
  companion to Andromeda. The left panel shows that there is only a
  very slight color gradient in this galaxy which begins only in its
  outer regions, if at all. Surface photometry of this galaxy is
  especially difficult given that it is projected on the complex
  background of spiral structure in M31. The right-hand panel shows
  the divergence of photometry based on earlier photographic
  photometry (circles) and the GALEX observations (triangles). }
\end{figure}

\subsection{M83: Extinction and Extended UV Disks}

\begin{figure}
  \includegraphics[height=.65\textheight]{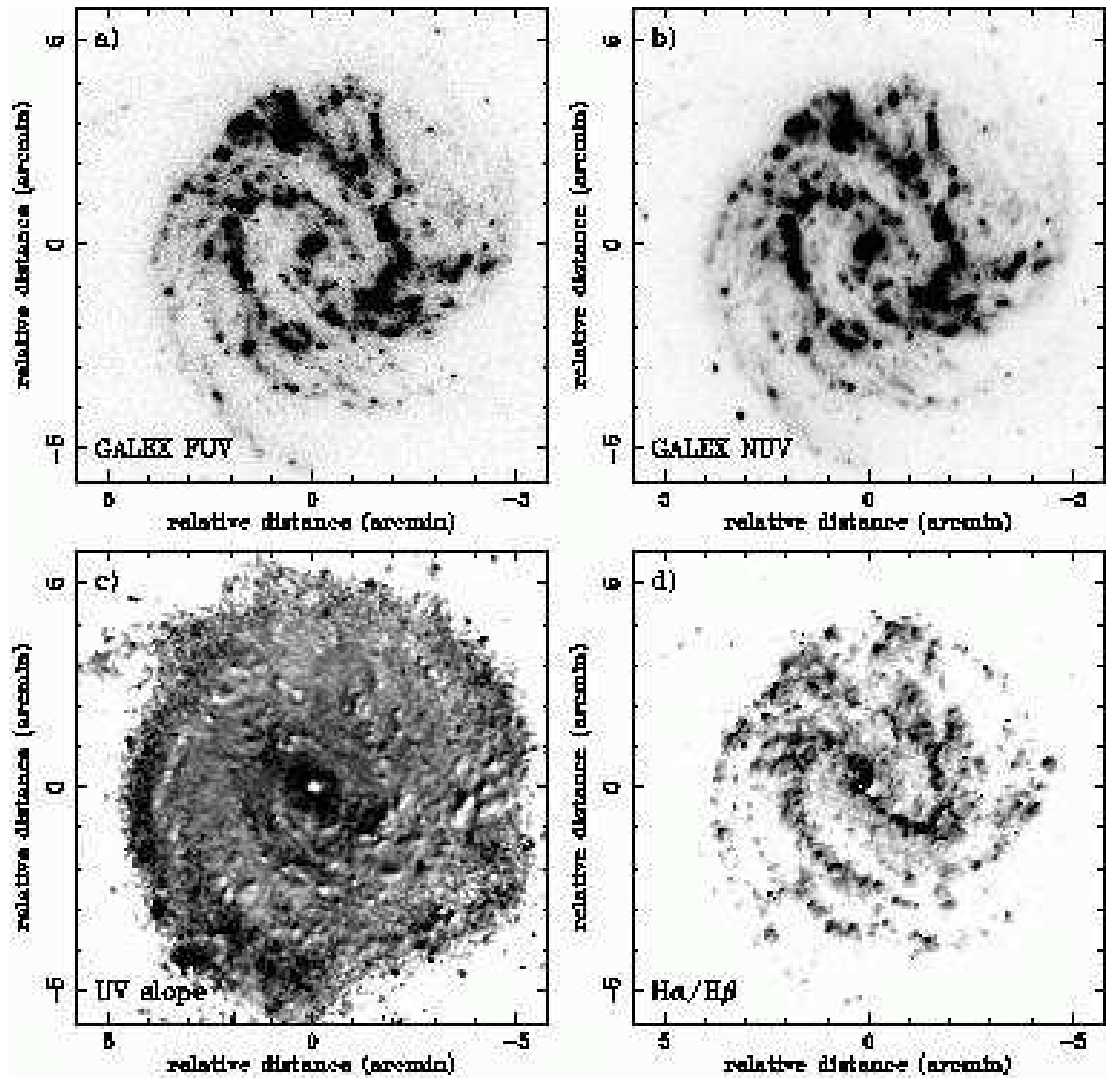}
  \caption{M83: (a) FUV image (from the sky background to 22.3 AB mag
  arcsec$^{-2}$), (b) NUV image (from the sky background to 22.7 AB
  mag arcsec$^{-2}$), (c) UV spectral slope, from $-$2.5 (white) to
  0.3 (black).  (d)\ha/\hb{} ratio, from \ha/\hb=2 to 6. White areas
  in this plot correspond to regions where either H$\alpha$ or
  H$\beta$ was below 3 $\sigma_{\mathrm{sky}}$. 8''$\times$8'' median
  filters were applied in the case of panels (c) \& (d). (Taken from
  Boissier {\it et al.}  \cite{boi05}}
\end{figure}

Boissier {\it et al.} \cite{boi05} used FUV and NUV images of M83 to compute
the radial profile of the UV spectral slope 
disk. They briefly present a model of its chemical evolution which
allows them to obtain realistic intrinsic properties of the stellar
populations.  Using corollary data, they also compute the profiles of
\ha/\hb{} and of the total IR (TIR) to FUV ratio.  Both data and model
are used to estimate and compare the extinction gradients at the FUV
wavelength obtained from these various indicators. They discuss the
implications for the determination of the star formation rate.

\begin{figure}
  \includegraphics[height=0.7\textheight]{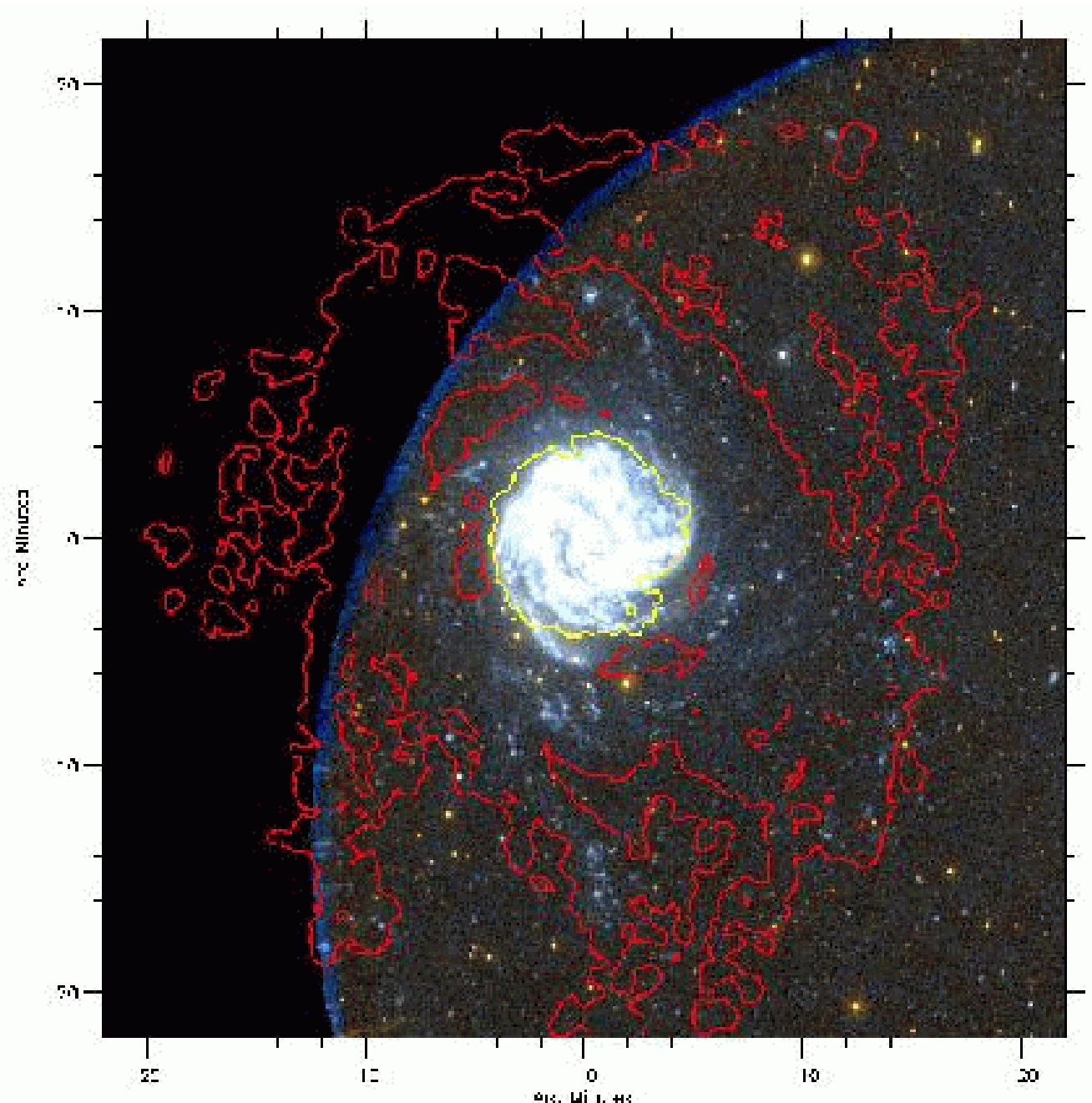} \caption{GALEX
  Imaging of M83. The grey-scale background image of M83 is derived
  from a combination of the FUV and NUV GALEX data. The contours
  correspond to the HI distribution published by Tilanus \& Allen
  (1993) extend out to the limits of the image (44~arcmin diameter,
  which corresponds to a physical extent of about 60~kpc at the
  distance of M83; the inner disk of M83 as defined by its inner
  high-surface-brightness disk HII regions is about 5 times
  smaller (from Thilker {\it et al.} \cite{thi05b}. }
\end{figure}

Thilker {\it et al.} \cite{thi05b} have discovered an extensive sample of
UV-bright stellar complexes in the extreme outer disk of M83 (Figure
6), extending four to five times the radius where the majority of
$H~II$ regions are detected ($R_{HII}$ = 5.1~arcmin or 6.6 kpc).
These sources are typically associated with large-scale filamentary
$H~I$ structures in the warped outer disk of M83, and are distributed
beyond the galactocentric radii at which molecular ISM has yet been
detected.  It is of interest that only a subset of the outer disk UV
sources have corresponding $H~II$ regions detected in H$\alpha$
imaging, consistent with a sample of mixed age in which a few sources
are a few Myr old and the majority is much more evolved ($\sim 10^8$
yr).

\subsection{M101: UV Leakage and Recent Star Formation}

A comparative study of the spatial distribution of UV and far-infrared
fluxes across the disk of M~101 has been undertaken by Popescu {\it et
al.} \cite{pop05}. By comparing total UV emission from both GALEX bands with
total far-infrared (FIR) flux from ISOPHOT (60, 100 and 170 $\mu$m)
they discovered a very strong correspondence of the FIR/UV ratio with
galactocentric radius (Figure 7). The ratio decreases monotonically
from values somewhat in excess of 3 near the nucleus to values near
zero in the outer regions of the galaxy. This they attribute to a
large-scale distribution of diffuse dust decreasing in optical depth
with radius and dominating over the more localized variations in
opacity occurring between the arm and interarm regions. They also find
a tendency for the FIR/UV ratio (at a given radius) to take on higher
values in the regions of diffuse interarm emission and lower values in
the spiral-arm regions. This is interpreted both in terms of the escape
probability of UV photons from spiral arms and their subsequent
scattering in the interarm regions, and in terms of the larger
relative contribution of optical photons to the heating of the dust in
the interarm regions.

\begin{figure}
  \includegraphics[height=0.4\textheight]{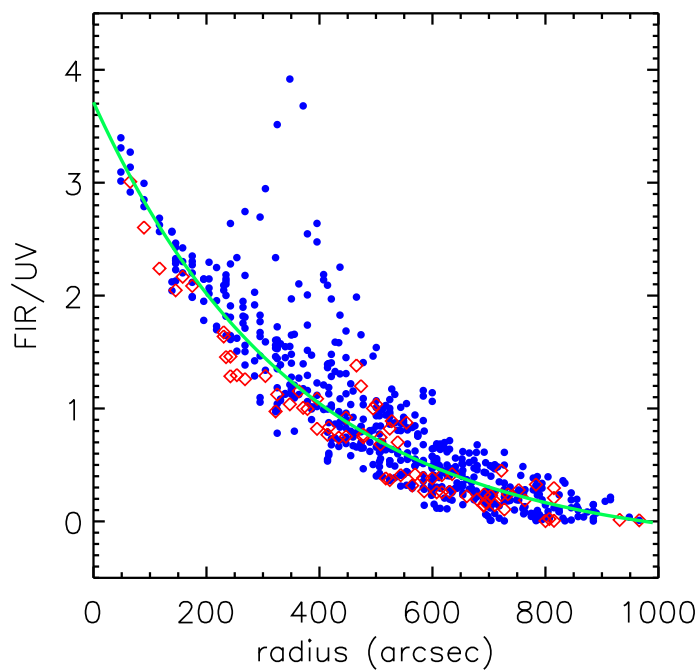} \caption{The pixel
  values of the FIR/UV ratio map of M101 at the resolution of the
  170\,${\mu}$m image versus the galactocentric radius. The
  dots are for lines of sight towards interarm regions and the
  diamonds are towards the spiral arm regions. The solid line is an
  offset exponential fit to the data. (Adapted from Popescu {\it et
  al.} \cite{pop05}}
\end{figure}

Another trend with galactocentric radius in M101 (and M51) is reported
and discussed by Bianchi {\it et al.} \cite{bia05b}. Using GALEX and SDSS
photometry of compact stellar clusters in M101 they have used
population synthesis models to derive ages, reddenings luminosities
and current/initial masses. A galactocentric gradient in the FUV/NUV
color index suggests younger clusters are preferentially found in the
outer parts of the disks, with M101 showing the more pronounced
gradient as compared to M51. The radial profiles in FUV/NUV for other
galaxies (M31, M33 and M83) show the same trend; however, the
magnitude of the effect is variable from galaxy to galaxy.

\subsection{Interacting Galaxies and Tidal tails}

Neff {\it et al.} \cite{nef05} report the detection of significant FUV and NUV
emission from stellar substructures within the tidal tails of four
on-going galaxy mergers: ARP 157, ARP 295, NGC 5713/19 and NGC
7769/71.  The UV-bright regions are optically faint and are coincident
with $HI$ density enhancements.  FUV emission is detected at any
location where the $HI$ surface density exceeds $\sim$ 2 M$_{\odot}$
pc$^{-2}$, and is often detected in the absence of any visible
wavelength emission. One example of UV tails and bridges, taken from
their paper, is shown in Figure 8.  UV luminosities of the brighter
regions of the tidal tails imply masses of 10$^6$M$_{\odot}$ up to
$\sim$ 10$^9$M$_{\odot}$ in young stars in the tails, and $HI$
luminosities imply similar $HI$ masses.  UV-optical colors of the
tidal tails indicate stellar populations as young as a few Myr, and in
all cases ages $<$ 400Myr.  Most of the young stars in the tails
formed in single bursts rather than resulting from continuous star
formation, and they formed {\it in situ} as the tails evolved.  Star
formation appears to be older near the parent galaxies and younger at
increasing distances from the parent galaxy.  The youngest stellar
concentrations, usually near the ends of long tidal tails, have masses
comparable to confirmed tidal dwarf galaxies and may be newly forming
galaxies undergoing their first burst of star formation.

A more detailed study of GALEX observations of an interacting galaxy,
the Antennae, has been undertaken by Hibbard {\it et al.}
\cite{hib05}. The tidal $UV$ morphology seen there is
remarkably similar to the tidal neutral hydrogen gas, with a close
correspondence between regions of bright $UV$ emission and high $HI$
column density. There are interesting features in the $UV$ morphology
of the inner regions. The $FUV$ light in the south-western half of the
disk of NGC 4039 is sharply truncated, coincident with a similar
truncation in the $HI$ disk. This gas may have been removed by the X-ray
loops recently imaged in this system by {\tt Chandra}, inhibiting
subsequent star formation. On a larger scale, there is a $FUV$
``halo'', possibly related to shocked gas in the X-ray halo.

While it has long been known that tails are generally blue, the GALEX
observation provide color baselines to determine how much of the $UV$
radiation comes from stars younger than the dynamical age of the
tails. The preliminary analysis by Hibbard {\it et al.} \cite{hib05}
suggests that most of the stars within the tidal tails pre-date the
tail formation period. The population in the northern tail appears
older than that in the southern tail. Examining individual $UV$-bright
regions within the tails they find regions of more recent star
formation, which occur at regions of higher $HI$ column density,
and extend well beyond the previously identified tail star forming
regions. Such tails are providing promising laboratories for the study
of star formation outside of the usual disk environment, and thereby
testing general theories of star and cluster formation.

\begin{figure}
  \includegraphics[height=0.4\textheight]{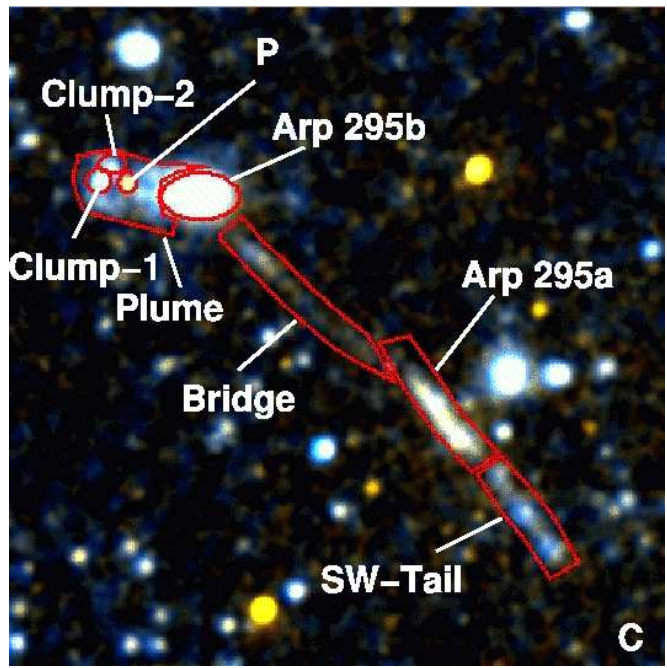} \caption{An example
of the tidal tails discussed in Neff {\it et al.} \cite{nef05}. In this case,
the interacting system Arp 295 nicely illustrates the UV clumps, tails and
bridges being found by GALEX in the outer regions of merging systems. }
\end{figure}

\vfill\eject
\begin{theacknowledgments}
I want to thank Cristina Popescu and Richard Tuffs for the opportunity
to report at their meeting on early results from the GALEX
mission. And I thank all of my colleagues on the GALEX science team
for their contributions to this interim review in specific and to the
overall scientific success of the mission to date. I have shamelessly
begged, borrowed and stolen images, plots, ideas and text from the
many papers so far published by the Team; I hope that proper
attribution is evident in this overview. A complete collection of
papers resulting from the first round of publishing by the GALEX Team
are slated to appear in a dedicated volume of the {\it Astrophysical
Journal (Letters)} early in 2005; and they can also be found in their
entirety at http://www.galex.caltech.edu/PUBLICATIONS/

The entire GALEX Team gratefully acknowledges NASA's support for
construction, operation, and science analysis for the GALEX mission,
developed in corporation with the Centre National d'Etudes Spatiales
of France and the Korean Ministry of Science and Technology. The
grism, imaging window, and uncoated aspheric corrector were supplied
by France .  We acknowledge the dedicated team of engineers,
technicians, and administrative staff from JPL/Caltech, Orbital
Sciences Corporation, University of California, Berkeley, Laboratory
Astrophysique Marseille, and the other institutions who made this
mission possible.  BFM was supported by the Observatories of Carnegie
Institution of Washington and by the NASA/IPAC Extragalactic Database
(NED).
\end{theacknowledgments}




\begin{thebibliography}{99}


\bibitem[\protect\citeauthoryear{Bianchi et al.}{2005a}]{bia05a} 
Bianchi, L., {\it et al.} 2005a, ApJ 619, L27

\bibitem[\protect\citeauthoryear{Bianchi et al.}{2005b}]{bia05b} 
Bianchi, L., {\it et al.} 2005b, ApJ 619, L71

\bibitem[\protect\citeauthoryear{Boissier et al.}{2005}]{boi05} 
Boissier, S., {\it et al.} 2005, ApJ 619, L83

\bibitem[\protect\citeauthoryear{Hibbard et al.}{2005}]{hib05} 
Hibbard, J., {\it et al.} 2005, ApJ 619, L87

\bibitem[\protect\citeauthoryear{Hoopes et al.}{2005}]{hoo05} 
Hoopes, C.G., {\it et al.} 2005, ApJ 619, L99

\bibitem[\protect\citeauthoryear{Martin et al.}{2005}]{mar05} 
Martin, D.C. {\it et al.} 2005, ApJ 619 L1

\bibitem[\protect\citeauthoryear{Morrissey et al.}{2005}]{mor05} 
Morrissey, P., {\it et al.} 2005, ApJ 619, L7

\bibitem[\protect\citeauthoryear{Neff et al.}{2005}]{nef05} 
Neff, S., {\it et al.} 2005, ApJ 619, L91

\bibitem[\protect\citeauthoryear{O'Connell et al.}{1992}]{con92} 
O'Connell, R. W., et al. 1992, ApJ, 395, L45

\bibitem[\protect\citeauthoryear{Ohl et al.}{1998}]{ohl98} 
Ohl, R. G., et al. 1998, ApJ, 505, L11

\bibitem[\protect\citeauthoryear{Popescu et al.}{2005}]{pop05} Popescu, C.C., 
{\it et al.} 2005, ApJ 619, L75

\bibitem[\protect\citeauthoryear{Gil de Paz et al.}{2005}]{gdp05} 
Gil de Paz, A., {\it et al.} 2005, ApJ 619, L115



\bibitem[\protect\citeauthoryear{Thilker et al.}{2005a}]{thi05a} 
Thilker, D.A., {\it et al.} 2005a, ApJ 619, L67

\bibitem[\protect\citeauthoryear{Thilker et al.}{2005b}]{thi05b} 
Thilker, D.A., {\it et al.} 2005b, ApJ 619, L79

\bibitem[\protect\citeauthoryear{Tilanius \& Allen}{1993}]{tilall93} 
Tilanius, R.P.J., \& Allen, R.J. 1993, A \& Ap, 274, 707





\end{thebibliography}
\end{document}